# Forecasting Shock-associated Energetic Particle Intensities in the Inner Heliosphere: A Proof-of-Concept Capability for the PUNCH Mission

M. A. Dayeh[1,2], M. J. Starkey[1], H. A. Elliott[1,2], R. Attie[4,5], C. E. DeForest[3], R. Bučík[1], and M. I. Desai[1,2]

[1] *Southwest Research Institute, San Antonio, TX 78238 (maldayeh@swri.edu)*

[2] *University of Texas at San Antonio, San Antonio, TX 78249*

[3] *Southwest Research Institute, Boulder, CO, 80302*

[4] *NASA Goddard Space Flight Center, MD, 20771*

[5] *George Mason University, VA, 22030*

**Abstract.**

Solar energetic particles (SEPs) associated with shocks driven by fast coronal mass ejections (CMEs) or shocks developed by corotating interaction regions (CIRs) often extend to high energies, and are thus key elements of space weather. The PUNCH mission, set to be launched in 2025, is equipped with photometric that enables 3D tracking of solar wind structures in the interplanetary space through polarized light. Tracking techniques are used to estimate speeds and speed gradients of solar structures, including speed jumps at fast shocks. We report on a strong and a robust relation between the shock speed jump magnitude at CME and CIR shocks and the peak fluxes of associated energetic particles from the analysis of 59 CME-driven shocks and 74 CIRs observed by Wind/STEP between 1997-2023. We demonstrate that this relation, along with PUNCH anticipated observations of solar structures can be used to forecast shock-associated particle events close to the Sun; thus, advancing and providing a crucial input to forecasting of SEP fluxes in the heliosphere.

## 1. Introduction

Energetic particles in interplanetary space play a critical role in space weather, posing significant challenges for deep space exploration (Whitman et al. 2022 and references there in). These particles originate from various sources within and beyond our heliosphere, spanning solar and galactic sources (Desai and Giacalone, 2016). Galactic cosmic rays (GCRs) undergo periodic changes over extended timeframes, offering space weather predictability and thus simplifying related risk mitigation strategies. In contrast, solar energetic particles (SEPs) encompass hard-to-predict elements, including sudden coronal eruptions, extreme enhancements of energetic particles, in addition to complex and intertwined mechanisms that continuously affect SEP transport and acceleration in the IP medium.

Primary populations of SEPs are known to be accelerated by (i) magnetic reconnection processes in flares (impulsive SEPs; e.g., Benz & Güdel 2010, Bučik et al. 2020), (ii) diffusive processes at shocks driven by fast coronal mass ejections (CMEs; e.g., Decker et al. 1981), or at compression regions where fast and slow solar wind speeds interact as they corotate with the sun, forming corotating interaction regions (CIRs; e.g., Mason et al. 2008). Furthermore, locally-trapped and accelerated material at CME-driven shocks comprise very high fluxes of relatively low energy particles (~up to tens of MeV/n), often observed superposed on ICME-associated SEP profiles and referred to as energetic storm particles (ESPs; e.g., Desai et al. 2003).

SEPs present notable radiation risks across various domains. Their impact extends to commercial aviation, influencing communication and navigation, particularly in polar flights. Additionally, astronauts exposed to this radiation in space face both immediate and prolonged health issues, including heightened cancer risks (Chancellor et al., 2014; Onorato et al., 2020). SEPs could also interfere with spacecraft electronics, resulting in operational disruptions and instrument degradation, data interference, or in severe cases, cause irreversible damages (Horne et al. 2013; Maurer et al. 2017).

CME and CIR shocks are associated with distinct plasma signatures as they sweep by spacecraft. These signatures are directly related to their shocks' properties, such as strength and geometry. Several studies have extensively investigated the properties of shocks, aiming to establish statistical relationships that could facilitate predictions for SEPs and associated particles, often reporting moderate correlations due to various reasons (e.g., Lee et al. 2012; Moreland et al. 2023). Among the quantities correlating with the energetic particle component, the parameter $\Delta V$

(difference between upstream and downstream solar wind speed) stands out as a strong parameter in ESPs (Dayeh et al. 2018), and similarly for CIRs (Bučík et al. 2009).

Accurate forecasts of SEP properties with long lead times are crucial for space weather operations. One possible way to extend lead time is to use remote solar, coronal, and heliospheric imaging observations to detect solar disturbances closer to the sun. A main hurdle of remote observations is to determine where along the line-of-sight a structure is located such that the full 3-D motion can be tracked. New strides are being made to solve this problem. One example is the upcoming Polarimeter to UNify the Corona and Heliosphere (PUNCH) mission (DeForest et al., 2022), which comprises a four-microsatellite constellation in Sun-synchronous low Earth orbit (LEO). PUNCH objective is to image the transition zone between the outer solar corona and the solar wind in the inner heliosphere from 6 solar radii (Rs) to 180 Rs (0.03 to 0.84 au) in polarized visible light. Polarized observations enable the full 3-D motion of structures to be determined and continuously tracked from the corona and into the heliosphere.

Although PUNCH is a science mission and the primary mission does not support low-latency data, PUNCH will allow scientific studies of shock imaging as means to forecast SEPs. PUNCH can estimate the shock speed jump in two ways: first, by direct tracking of small inhomogeneities that converge on a shock front, through the "J-map" technique that has been used to identify and understand CIR signatures using imaging data from STEREO/SECCHI and SMEI (Tappin & Howard 2009); and second, by direct photometric estimation of the density ratio on opposite sides of well-presented shocks (in which the line of sight is tangent to the shock face). Shock density ratio is directly related to the shock speed jump. Measuring the total brightness of a shock front compared to the "background corona" requires only a single image of the shock and is a proxy for longer-term measurements of the speed jump. This can be done by tracking individual small-scale blobs (e.g. Sheeley et al. 1997), which requires analyzing several hours of data. The two measurements are independent and can be used to cross-check plausibility of the inferred shock characteristics. That said, we note that not all these shocks necessarily affect the near-Earth environment from a space weather perspective. The variability of shock parent location and geometry (e.g., quasi-parallel or -perpendicular) also affect the localized accelerated particle profile and significantly impact the observed particle intensities. However, we emphasize that this work focuses on providing a potential to advance shock-associated SEP forecasting by including PUNCH tracking capabilities.

In this study, we examine the properties of 59 CME-driven shocks and 74 CIR shocks associated with clear particle enhancements. We find a strong correlation between the energetic particle peak fluxes and the speed jump, ΔV, at ICME and CIR driven shocks. Therefore, imaging measurements such as those from PUNCH may enable long lead time forecasting and quantification of energetic particles peak fluxes from CME-driven shocks and CIR compression regions.

## 2. Data Analysis and Methodology

Measurements of ESP and CIR associated energetic particles are obtained from the STEP instrument within the Energetic Particles, Acceleration, Composition, and Transport Experiment (EPACT; von Rosenvinge et al. 1995) onboard the Wind spacecraft. We utilize the Helsinki shock database (http://ipshocks.fi), an ICME list (Richardson and Cane 2010), and a CIR list (Broiles et al. 2013) as a starting point for the analysis. Figure 1a and Figure 1b show temporal profiles of He intensities for an ESP event and a CIR event respectively, along with the solar wind speed showing the speed jump in association with the shock arrival (Figure 1c and Figure 1d). For ESP events, we impose the following selection requirements: (i) an identified CME must be present within a 12-hour window after the shock time; (ii) no ICMEs are identified within 24 hours before the shock; (iii) a quiet interplanetary pre-event period before the shock; and (iv) $avg(V_{sw}) <$ 500 $km/s$ to insure the ICME is not in a fast solar wind stream. The third requirement is enforced by constraining the coefficient-of-variation statistic ($\sigma_m$; defined as the standard deviation divided by the mean) of the density, solar wind speed, and magnetic field strength within 36-hour window before shock time ($\sigma_{m,V} < 0.15$, $\sigma_{m,n} < 0.38$, and $\sigma_{m,B} < 0.4$). These numbers were chosen to optimize the number of events used and maintain the intended quiet conditions.

For CIR events, we only require the event to have an energetic particle enhancement at energies 0.1 – 0.5 MeV in association with the reverse boundary (RB) shock passage. A flux increase is determined by requiring the maximum integrated flux within a 3-hour time window centered on the RB to exceed the average integrated flux,

$$F_{avg} = \frac{\sum_{390\ keV}^{150\ keV}[\Delta E_i \cdot F(E_i)]}{\sum_{390\ keV}^{150\ keV}[\Delta E_i]}$$

within this window. $F_i$ is the differential flux within the energy pass band $\Delta E_i$. These selection criteria result in 59 ESP events and 74 CIR events.

Figure 1e and Figure 1f show the observed He peak flux at different energies associated with the shocks for ESP and CIR events as a function of solar wind speed jump. The Pearson correlation coefficient is shown along with the weighted fit analytic form. Statistical testing shows that the correlation in both cases is statistically significant. As a sensitivity check, we have also performed a "leave-one-out cross-validation" (LOOCV; e.g., Hastie et al. 2009) and found that the correlation coefficient only varies by a maximum of 8% for CIRs (4% for ESPs) if we remove any of the events, indicating that single events do not drive the inferred correlation (Figure 1g and Figure 1h). It is important to note that the exponent of the power-law relationship between the peak flux and dV defines the level of uncertainty of the predicted peak flux. The small exponent means that an error in the speed measurement would produce a smaller error in the peak flux. Furthermore, while the correlation does not imply precise predictability for individual events given the large spread of peak fluxes at any given shock speed jump, it does offer a probabilistic forecast window, allowing us to estimate an intensity range for given shock speeds. We present an analytical form to establish a benchmark, should the fitted equation be needed for future applications."

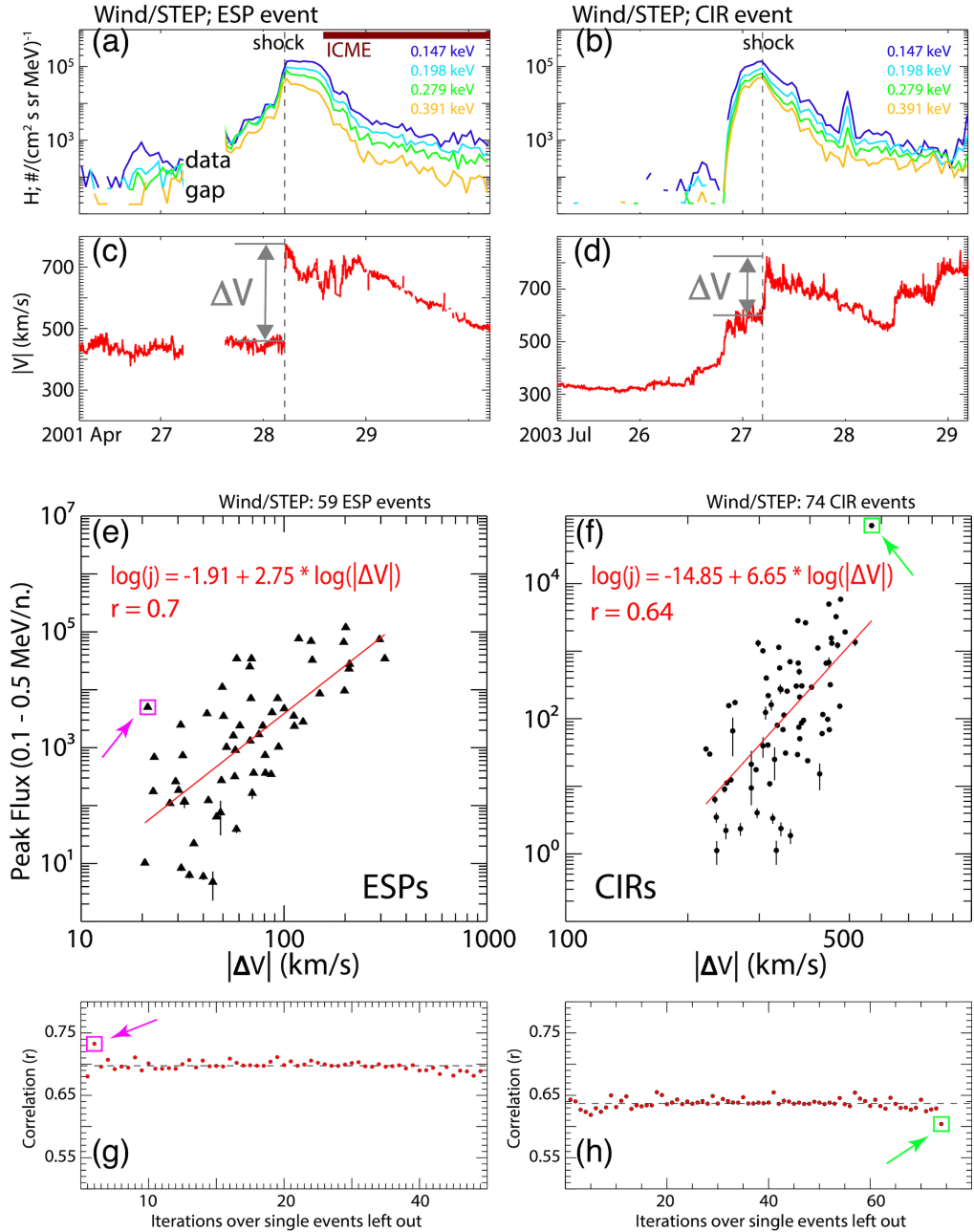

***Figure 1.*** *(a) Temporal profiles of H intensities between 0.1-0.5 MeV in an ESP and a CIR (b) event, along with the solar wind profiles (c,d). (e,f) Scattered plots showing the correlation and the weighted fits between observed peak fluxes and the speed jump in 54 ESP events and 74 CIRs. Bottom panels demonstrate how the correlation varies if one event is removed, and colored boxes and arrows indicate the events affecting the correlation most (see text for details). The correlation is very stable and is not affected by a single measurement.*

Figure 2a shows an illustration of an advanced flow tracking method currently being developed further by the PUNCH team. This method offers a high degree of granularity when tracking density structures propagating in the corona. This method is based on the so-called "Magnetic Balltracking" method, originally developed by Attie & Innes (2015) for tracking moving magnetic fragments in magnetograms. In coronal imagery, the algorithm is being adapted and tested using the imagery of the deep exposure campaigns of STEREO/COR2 (Deforest et al. 2018). With minor modification of the algorithm, it detects, tracks and labels coherent density structures, including CMEs, from which we can derive the mean radial velocity (Figure 2b) around the sun at a few degrees of azimuthal resolution. On COR2 imagery, the accuracy of the mean radial velocity is estimated to range from ~5% to ~15% in 10-degree bins, depending on the time averaging window and contrast of the density structure. We expect a similar level of accuracy with PUNCH imagery.

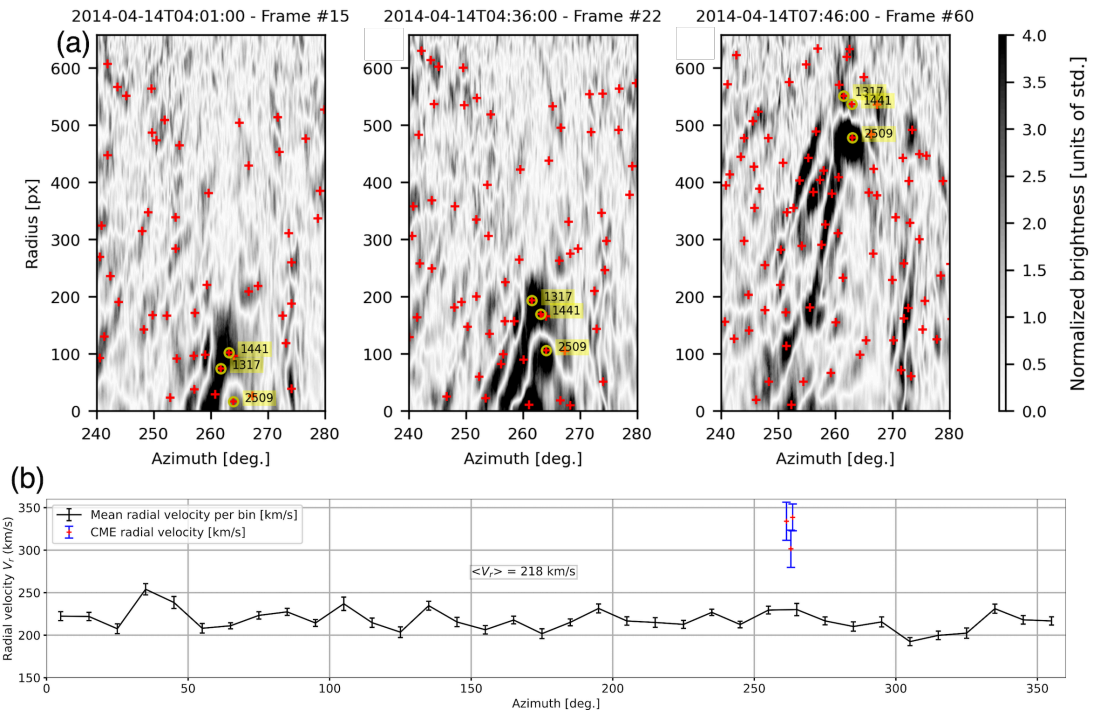

***Figure 2:*** *(a) Example of the tracking of plasma density structures with Magnetic Balltracking. The red crosses are at the location of the balls that track the density structures. We also show an*

*example of the detection and tracking of a CME's leading and trailing brightness patterns tagged in yellow. The scale of the y-axis is 1px ~9.36 solar radii ~ 10170 km. (b) Mean radial velocities as a function of the azimuthal position around the sun, estimated by averaging the motion of the balls of Magnetic Balltracking over ~6.6 hr. The red cross and blue error bars show the average velocity of three density structures tracked in a CME (yellow tags in (a).*

## 3. Discussion and Implications

Analysis from a large set of ESPs and CIRs associated with particle enhancements show that there is a strong relation between the peak intensities and speed jump across the shock. Coronagraphs and heliospheric imaging can be used to track CMEs as they move away from the Sun. The leading edges of fast CMEs is where the interplanetary shocks are located and where there is a sharp boundary between the background wind and the fast CME running into the background. By tracking the speed of these leading edges/shocks headed toward Earth, the shock speed is determined, and shock Earth arrival time estimated. The speed difference between the leading edge/shock and the background wind is proportional to the speed jump of the associated moving shock. The relation between the peak intensities and the speed jump shown in Figure 1b,c can thus be utilized to provide a reliable estimate of the peak intensities at the observed energies arriving at Earth in association with the local particles.

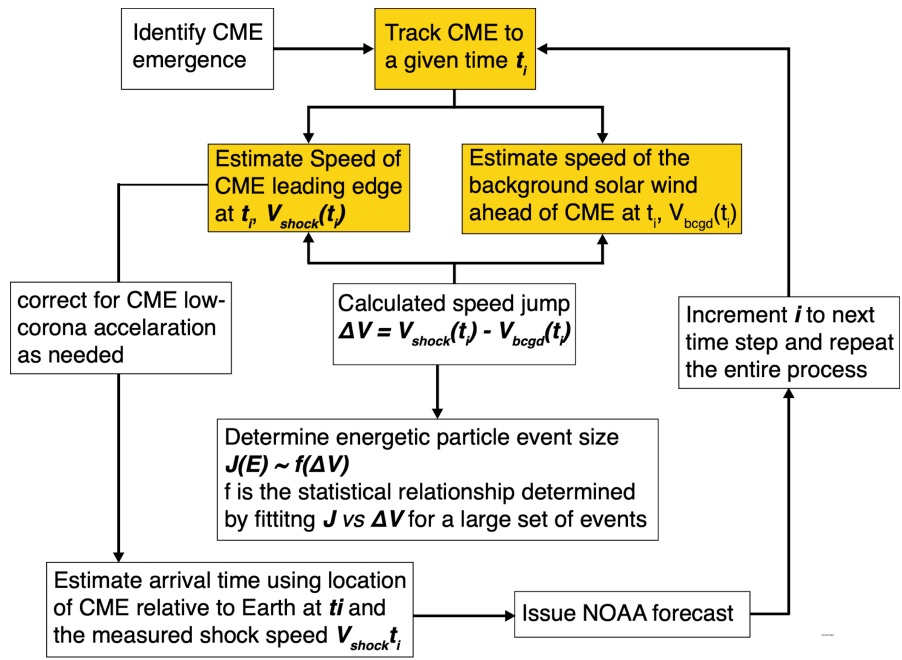

***Figure 3**: Flow diagram illustrating the steps needed to produce real-time forecasts of ESPs using a real-time coronagraph and heliospheric imaging pipeline.*

In Figure 3, we provide a potential flowchart of how this observation could be pipelined for a space weather forecasting application. Here, we outline a process for performing real-time forecasting of ESP and SEP-associated CIR events. The shock/leading edge speed, and speed jump is determined by tracking ICMEs and CIRs in coronagraph and heliospheric images. By estimating the leading-edge speed relative to the background, wind speed the speed jump at the shock can be estimated. Because the speed jump correlates well with the peak ESP and CIR-associated SEP flux, a dynamic forecast can then be provided in real-time based on how the shock strength evolves and how the leading edge of the ICME/CIR gets forecasted. As the CME gets closer to Earth, new images can be used to create new updated forecasts of $\Delta V$ which are then used to produce new updated and more accurate forecasts of the associated peak fluxes of associated energetic particles.

In this paper, we provide observational evidence that PUNCH measurements could play an important factor in increasing the reliability of forecasting the occurrence and properties of shock-accelerated particles as shocks travel through the interplanetary space and near-Earth environment.

## 4. Acknowledgements

This work was partially funded through PUNCH, a NASA Small Explorer mission, via NASA Contract No. 80GSFC18C0014. Partial support came from NASA LWS grants 80NSSC19K0079, 80NSSC21K1316, *80NSSC21K1307*, 80NSSC20K1815, 80NSSC24K0908; PSP GI award 80NSSC21K1769; and O2R 80NSSC21K0027.